\documentclass[sigconf]{acmart}
\AtBeginDocument{%
  }

\copyrightyear{2026}
\acmYear{2026}
\setcopyright{cc}
\setcctype{by-nc-nd}
\acmConference[SIGIR '26] {Proceedings of the 49th International ACM SIGIR Conference on Research and Development in Information Retrieval}{July 20--24, 2026}{Melbourne, VIC, Australia.}
\acmBooktitle{Proceedings of the 49th International ACM SIGIR Conference on Research and Development in Information Retrieval (SIGIR '26), July 20--24, 2026, Melbourne, VIC, Australia}
\acmISBN{979-8-4007-2599-9/2026/07}
\acmDOI{10.1145/3805712.3809849}

\usepackage{float}
\begin{document}

\title{Counterfactual Multi-task Learning for Delayed Conversion Modeling in E-commerce Sales Pre-Promotion}

\author{Xin Song}
\affiliation{%
  \institution{Alibaba Group}
  \city{Beijing}
  \country{China}
}
\email{xin_xiaoye@bupt.cn}

\author{Kaiyuan Li}
\affiliation{%
  \institution{Alibaba Group}
  \city{Beijing}
  \country{China}}
\email{tsotfsk@bupt.cn }

\author{Jinxin Hu}
\affiliation{%
  \institution{Alibaba Group}
  \city{Beijing}
  \country{China}
}
\email{jinxin.hjx@alibaba-inc.com}
\authornote{Corresponding author.}


\begin{abstract}
Sales promotions, as short-term incentives to stimulate product purchases, play a pivotal role in modern e-commerce marketing strategies. During promotional events, user behavior patterns exhibit distinct characteristics compared to regular periods. Notably, in the pre-promotion phase, users typically engage in product search and browsing without immediate purchases, often adding items to carts in anticipation of promotional discounts. This behavior leads to delayed conversions, resulting in significantly lower conversion rates (CVR) before the promotion day.

Although existing research has made progress in CVR prediction for promotion days using historical data, it largely overlooks the critical pre-promotion period. Although delayed feedback modeling has been extensively studied, current approaches fail to account for the unique distribution shifts in conversion behavior before promotional events, where delayed conversions predominantly occur on the promotion day rather than over continuous time windows.

To address these limitations, we propose the Counterfactual Multi-task Delayed Conversion Model (CM-DCM), which leverages historical pre-promotion data to enhance CVR prediction for both delayed and direct conversions. Our model incorporates three key innovations: (i) A multi-task architecture that jointly models direct and delayed conversions using historical pre-promotion data; (ii) A personalized user behavior gating module to mitigate data sparsity issues during brief pre-promotion periods; (iii) A counterfactual causal approach to model the transition probability from add-to-cart (ATC) to delayed conversion.

Extensive experiments demonstrate that CM-DCM outperforms state-of-the-art delayed CVR models in pre-promotion scenarios. Online A/B tests during major promotional events showed significant improvements in advertising revenue, delayed conversion GMV, and overall GMV, validating the effectiveness of our approach.
\end{abstract}



\maketitle

\section{Introduction}
\label{sec:intro_related}

Sales promotion refers to marketing campaigns launched by e-commerce companies to increase user interest or demand for their product or service~\cite{effects_promption, inf_prootion, post-promotion}. It is often held together with festivals (e.g., Christmas sales, Thanksgiving deals) or on specific dates such as Amazon Black Friday and Alibaba Double 11. On promotion day, products are offered at steep discounts, attracting millions of users and generating a substantial fraction of annual Gross Merchandise Volume (GMV) and orders. This makes accurate CVR prediction during the promotion cycle critical for platform revenue and advertisements delivery efficiency.

\begin{figure}[h]
  \centering
  \includegraphics[width=1.0\linewidth]{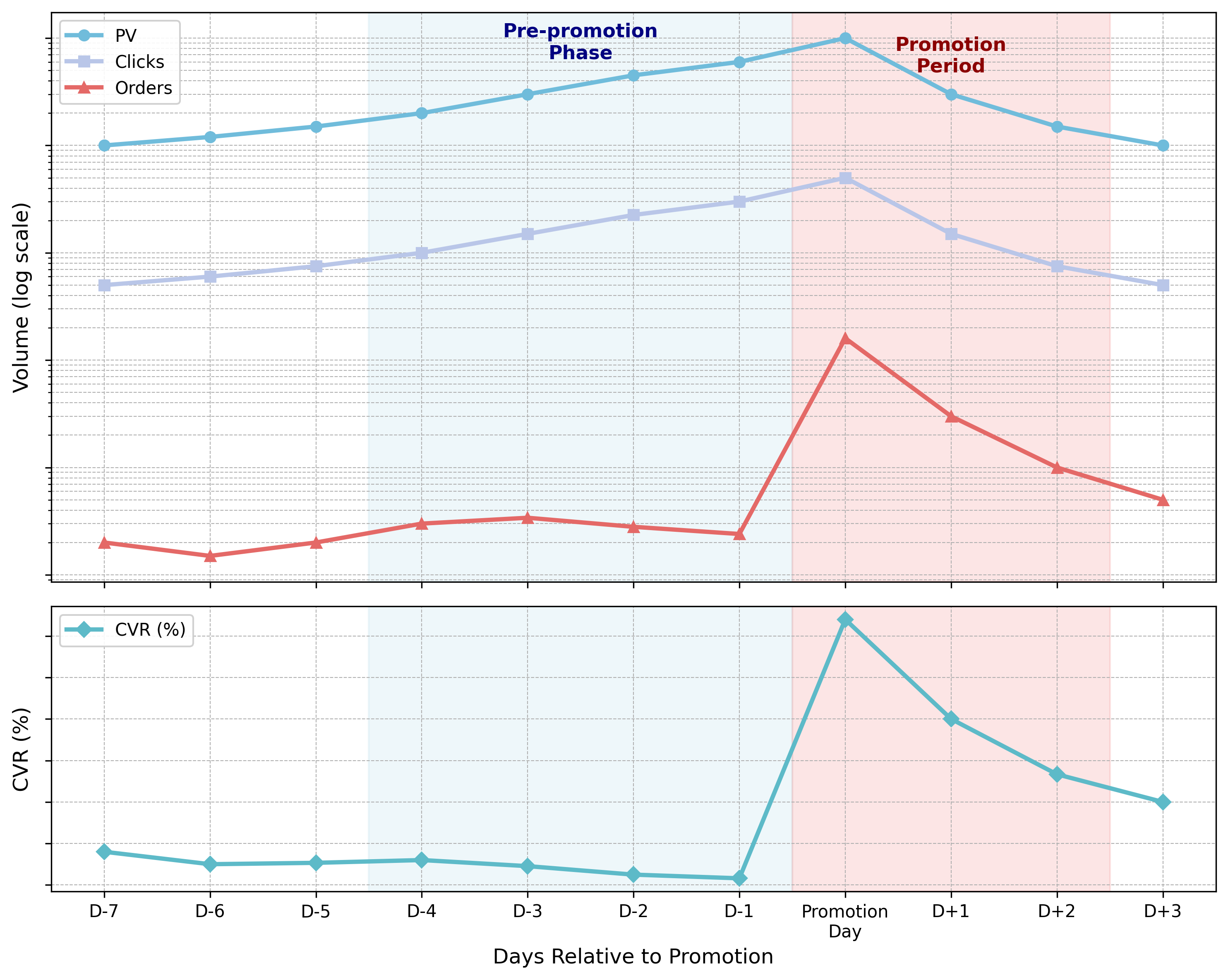}
  \caption{User behavior trend during a Sales Promotion.}
  \label{fig:1}
\end{figure}

User behavior shifts dramatically before the promotion day. During this \textit{pre-promotion} period—starting when the event is announced—users intensively browse and add items to cart but deliberately delay purchases until the promotion day. As shown in Figure~\ref{fig:1}, CVR drops sharply in pre-promotion, yet about 45\% of eventual promotion-day orders originate from pre-promotion clicks. This creates a unique delayed feedback problem: conversions are not uniformly delayed but concentrated on a single future day, violating assumptions of standard delayed CVR models.

Compared with well-defined click-through rate(CTR) prediction, two core challenges arise: (1) an \textit{out-of-distribution} (OOD) shift, as daily-trained CVR models fail to capture pre-promotion intent; and (2) a \textit{skewed delay distribution}, where conventional delayed feedback methods~\cite{DFM, FNW, ESDFM, DEFER, DEFUSE}—which assume stable conversion behavior and continuous delay windows—perform poorly. While Chan et al.~\cite{HDR} addressed CVR underestimation \textit{on} promotion day via historical date reuse, their method does not target the pre-promotion phase, where behavior patterns and label delays are fundamentally different.

In e-commerce platforms (e.g., Amazon, Taobao), the shopping cart (Add-to-Cart, ATC) acts as a temporary holding space for items users plan to purchase, bridging browsing and checkout. ATC behavior is a strong indicator of purchase interest, as it reflects explicit user intent—items added to carts are far more likely to convert than those merely viewed (e.g., 3-5× higher conversion rates). The timing and context of ATC (e.g., during promotions) further reveal urgency and price sensitivity, making it a critical signal for recommendation systems. A key insight is that the add-to-cart (ATC) action in pre-promotion is not merely correlated with—but likely \textit{causally influences}—delayed conversion. However, observational ATC–conversion correlations are confounded by factors like user intent or item popularity. This motivates a causal inference approach. Recent CVR works have begun incorporating causal reasoning to mitigate selection bias and confounding~\cite{ESMM, ESCM2, ESMMBD, ModelingTR}, yet none explicitly model the causal effect of pre-promotion ATC on promotion-day conversion under severe label delay.

To bridge this gap, we propose the \textbf{Counterfactual Multi-task Learning for Delayed Conversion Model} (CM-DCM). Our framework integrates: (i) a multi-task architecture that jointly models direct and delayed conversions using historical pre-promotion data; (ii) a counterfactual causal regularizer based on the Doubly Robust (DR) estimator to enforce the causal effect of ATC on delayed conversion; and (iii) a personalized gated transfer mechanism—inspired by PEPNet~\cite{PEPNet}—that adaptively transfers representations from pre-trained daily CVR/ATC models based on real-time user behavior. This enables robust, causally grounded CVR estimation in the data-scarce pre-promotion window, significantly improving ad targeting and platform GMV. To Summarize, the main contributions of our work are as follows:
\begin{itemize}
\item We propose a method named CM-DCM to model delayed conversions during the pre-promotion period by incorporating historical pre-promotion data and conversion labels from previous pre-promotions. To the best of our knowledge, this is the first research for delayed conversion modeling for the sales pre-promotion phase.
\item We further optimize the model's performance from multiple perspectives: multi-task modeling of delayed and direct conversions, personalized gated transfer for delay conversion modeling, counterfactual causal regularizer for ATC causal effect on delayed conversion.
\item We perform offline evaluations on public datasets and internal industrial datasets, and conduct online A/B tests on our online e-commercial advertising platform to demonstrate the effectiveness of our approach.
\end{itemize}

\section{Preliminaries}
\label{sec:preliminaries}
In e-commerce sales promotions (e.g., \textit{Double 11}, short for D11), user behavior during the \textbf{pre-promotion phase}—the days immediately preceding the main event—exhibits a distinct pattern: users actively browse and add items to their carts but deliberately \textbf{postpone their purchases} until the promotion day to capitalize on discounts. This leads to a significant portion of conversions being \textbf{delayed}, i.e., the conversion event occurs on the promotion day despite the click happening days earlier.

Formally, let $\mathcal{D}$ denote the set of all user-item click events during the pre-promotion period. For each event $i \in \mathcal{D}$, we observe a feature vector $\mathbf{x}_i \in \mathbb{R}^d$ that includes user profiles, item attributes, context, and real-time behavior sequences (e.g., add-to-cart). The ground-truth labels are defined as follows:
\begin{itemize}
\item $y_i^{\text{all}} \in \{0, 1\}$: the \textbf{all-conversion label}, which is 1 if the user converts (purchases) on the same day or the promotion day, and 0 otherwise.
\item $y_i^{\text{delay}} \in \{0, 1\}$: the \textbf{delay-conversion label}, which is 1 \textbf{only if} conversion occurs on the promotion day and 0 otherwise.
\end{itemize}
The relationship between these labels is summarized in Table~\ref{tab:label_def}. The primary goal of our task is to accurately estimate the \textbf{delayed conversion rate} (Delayed CVR)—the probability that a pre-promotion click will result in a purchase on the promotion day:
\begin{equation}
    p_i^{\text{delay}} = P(y_i^{\text{delay}} = 1 \mid \mathbf{x}_i)
    \label{eq:delay_cvr_def}
\end{equation}

\begin{table}[t]
\centering
\caption{Conversion types for a clicked item during pre-promotion and their label values.}
\label{tab:label_def}
\begin{tabular}{lcc}
\toprule
Conversion Type & $y^{\text{all}}_{\text{cvr}}$ & $y^{\text{delay}}_{\text{cvr}}$ \\
\midrule
Direct conversion & 1 & 0 \\
Delayed conversion & 1 & 1 \\
Non-conversion & 0 & 0 \\
\bottomrule
\end{tabular}
\end{table}

\section{Method}

\label{sec:method}

In this section, we present our proposed \textbf{Counterfactual Multi-task Learning for Delayed Conversion Model} (CM-DCM), a principled framework for delayed conversion rate (CVR) estimation in the pre-promotion phase of e-commerce sales events. CM-DCM addresses the unique challenges of this scenario—namely, severe label delay and distribution shift—by unifying three key components: (1) a multi-task learning architecture for joint modeling of direct and delayed conversions, (2) a personalized gating mechanism for effective knowledge transfer from pre-trained daily CVR/ATC models, and (3) a counterfactual causal regularizer to enforce the causal effect of the add-to-cart (ATC) action on delayed conversion.

\subsection{Multi-Task Learning Architecture for Delayed CVR}
\label{sec:pact}
\begin{figure}[t]
    \centering
    \includegraphics[width=\linewidth]{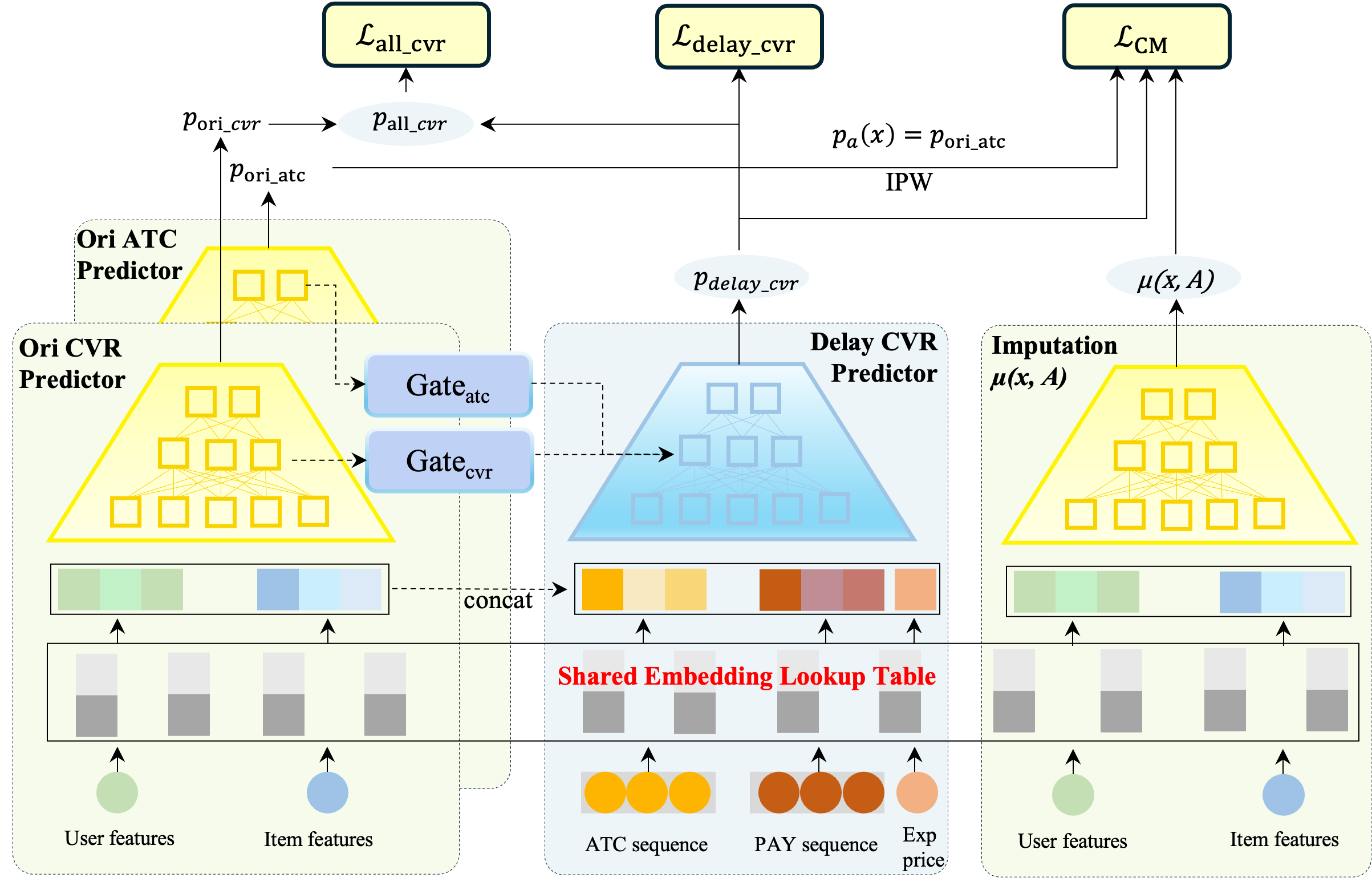}
    \caption{The architecture of the Counterfactual Multi-task Learning for Delayed Conversion Model (CM-DCM).}
    \label{fig:architecture}
\end{figure}

 The backbone of our regular CVR model shares the same structure as our former CTR models \cite{LREA,soft_ret}. As illustrated in Figure~\ref{fig:architecture}. The first tower, \textbf{Ori CVR Predictor} and \textbf{Ori ATC Predictor}, is a pre-trained multi-task model on daily data that predicts the immediate conversion probability and add-to-cart probability, denoted as $ p^{\text{ori}}_{\text{cvr}} $ and $ p^{\text{ori}}_{\text{atc}} $. The second tower, \textbf{Delay CVR Predictor}, is our primary module, which estimates the residual probability of a delayed conversion on the promotion day, denoted as $ p^{\text{delay}}_{\text{cvr}} $.

The final predicted probability for a user to convert (either directly or with delay) is the sum of these two components:
\begin{equation}
    p^{\text{all}}_{\text{cvr}} = \llbracket p^{\text{ori}}_{\text{cvr}} \rrbracket + p^{\text{delay}}_{\text{cvr}}
    \label{eq:all_cvr}
\end{equation}
where $ \llbracket \cdot \rrbracket $ denotes the stop-gradient operator, which freezes the parameters of the pre-trained tower during the fine-tuning phase to prevent its degradation on the daily conversion task.

The input to the Delay CVR tower is a rich feature vector that includes the frozen output of the last hidden layer from the Ori CVR tower, $ \llbracket \mathbf{h}^{L} \rrbracket $, An embedding of the item’s exposure price and discount ratio, $ \mathbf{E}_{\text{exp\_price}} $ and pooled representations of the user’s real-time behavior sequences for add-to-cart and purchase actions, $ \mathbf{V}_{\text{urb}}^{\text{atc}} $ and $ \mathbf{V}_{\text{urb}}^{\text{pay}} $.

Formally, the input is:
\begin{equation}
    \mathbf{input}_{\text{delay}} = \llbracket \mathbf{h}^{L} \rrbracket \parallel \mathbf{E}_{\text{exp\_price}} \parallel \mathbf{V}_{\text{urb}}^{\text{atc}} \parallel \mathbf{V}_{\text{urb}}^{\text{pay}}
    \label{eq:delay_input}
\end{equation}
where $ \parallel $ denotes vector concatenation. This input is passed through a multi-layer perceptron (MLP) to yield the delayed CVR:
\begin{equation}
    p^{\text{delay}}_{\text{cvr}} = \sigma(\text{MLP}_{\text{delay}}(\mathbf{input}_{\text{delay}}))
    \label{eq:delay_cvr_pred}
\end{equation}

The model is trained with a multi-task loss that jointly optimizes for the delayed conversion label $ y^{\text{delay}}_{\text{cvr}} $ and the all-conversion label $ y^{\text{all}}_{\text{cvr}} $:
\begin{equation}
    \mathcal{L} = \mathcal{L}_{\text{delay\_cvr}} + \lambda \mathcal{L}_{\text{all\_cvr}}
    \label{eq:mtl_loss}
\end{equation}
where $ \mathcal{L}_{\text{delay\_cvr}} $ and $ \mathcal{L}_{\text{all\_cvr}} $ are binary cross-entropy losses, and $ \lambda $ is a hyperparameter.

\subsection{Personalized Gated Representation Transfer}

To effectively leverage the rich user-behavior representations learned by the pre-trained daily CVR and ATC models, we employ a \textbf{personalized gated transfer} mechanism, inspired by PEPNet~\cite{chang2023pepnet}.

For each hidden layer $ i $ of the pre-trained towers, we extract their outputs $ \mathbf{h}^{i}_{\text{cvr}} $ and $ \mathbf{h}^{i}_{\text{atc}} $. Instead of directly concatenating these vectors, we modulate them with personalized gating signals that are a function of the user’s current pre-promotion behavior.

The gating networks are defined as:
\begin{align}
    \mathbf{g}^{i}_{\text{cvr}} &= \sigma(\text{Gate}^{i}_{\text{cvr}}([\mathbf{E}_{\text{user}}] \parallel \mathbf{V}_{\text{urb}}^{\text{atc}} \parallel \mathbf{V}_{\text{urb}}^{\text{pay}})) \\
    \mathbf{g}^{i}_{\text{atc}} &= \sigma(\text{Gate}^{i}_{\text{atc}}([\mathbf{E}_{\text{user}}] \parallel \mathbf{V}_{\text{urb}}^{\text{atc}} \parallel \mathbf{V}_{\text{urb}}^{\text{pay}}))
\end{align}
where $ \text{Gate}^{i}_{\text{cvr}} $ and $ \text{Gate}^{i}_{\text{atc}} $ are small MLPs, and $ \sigma $ is the sigmoid function.

The gated representations are then used as input to the corresponding layer of the Delay CVR tower:
\begin{equation}
    \mathbf{input}^{i}_{\text{delay}} = (\mathbf{g}^{i}_{\text{cvr}} \otimes \llbracket \mathbf{h}^{i}_{\text{cvr}} \rrbracket) \parallel (\mathbf{g}^{i}_{\text{atc}} \otimes \llbracket \mathbf{h}^{i}_{\text{atc}} \rrbracket) \parallel \ldots
    \label{eq:gated_input}
\end{equation}
where $ \otimes $ denotes element-wise multiplication. The personalized gating mechanism enables the model to account for product-specific consideration periods, dynamically adjusting information transfer based on both user activity and item characteristics. This ensures the model produces predictions that are more tailored to each product's context during the promotion timeline.

\subsection{Counterfactual Causal Regularization for ATC Effect}

A core insight is that the ATC action in the pre-promotion phase is a strong causal signal for a subsequent delayed conversion. To inject this causal prior into our model, we introduce a \textbf{counterfactual causal regularizer} based on the Doubly Robust (DR) estimator.

Let $ A \in \{0, 1\} $ be the ATC indicator, and $ Y \in \{0, 1\} $ be the delayed conversion label. Our goal is to estimate the individual causal effect (ICE) of ATC on $ Y $. We define an imputation model $ \mu(\mathbf{x}, A) $ that predicts the conversion outcome given the features $ \mathbf{x} $ and the ATC action. The DR estimator for the ICE is:
\begin{equation}
    \hat{\tau}_{\text{DR}}(\mathbf{x}) = \underbrace{[\mu(\mathbf{x}, 1) - \mu(\mathbf{x}, 0)]}_{\text{Outcome Regression}} + \underbrace{\frac{A(Y - \mu(\mathbf{x}, 1))}{p_a(\mathbf{x})} - \frac{(1-A)(Y - \mu(\mathbf{x}, 0))}{1 - p_a(\mathbf{x})}}_{\text{IPW Correction}}
    \label{eq:dr_ice}
\end{equation}
where $ p_a(\mathbf{x}) $ is the propensity score, which in our context is the predicted ATC probability from a pre-trained ATC model.

To integrate this into our main model, we use the DR estimate to construct a regularization term that encourages the main CVR prediction to align with the causal effect. Specifically, we add the following term to our loss:
\begin{equation}
    \mathcal{L}_{\text{CM}} = \mathbb{E} \left[ \left( p^{\text{delay}}_{\text{cvr}} - \mu(\mathbf{x}, 1) \right)^2 \right]
    \label{eq:cm_dcm_reg}
\end{equation}
This ensures that our model’s prediction for a user who has added an item to their cart is grounded in a counterfactually sound estimate of their delayed conversion potential.
The final loss function of CM-DCM combines all components:
\begin{equation}
    \mathcal{L}_{\text{CM-DCM}} = \mathcal{L}_{\text{delay\_cvr}} + \lambda \mathcal{L}_{\text{all\_cvr}} + \lambda_c \mathcal{L}_{\text{CM}}
    \label{eq:total_loss}
\end{equation}

  \begin{figure*}[h]
      \centering
      \begin{minipage}[b]{0.48\textwidth}
          \centering
          \includegraphics[width=\linewidth]{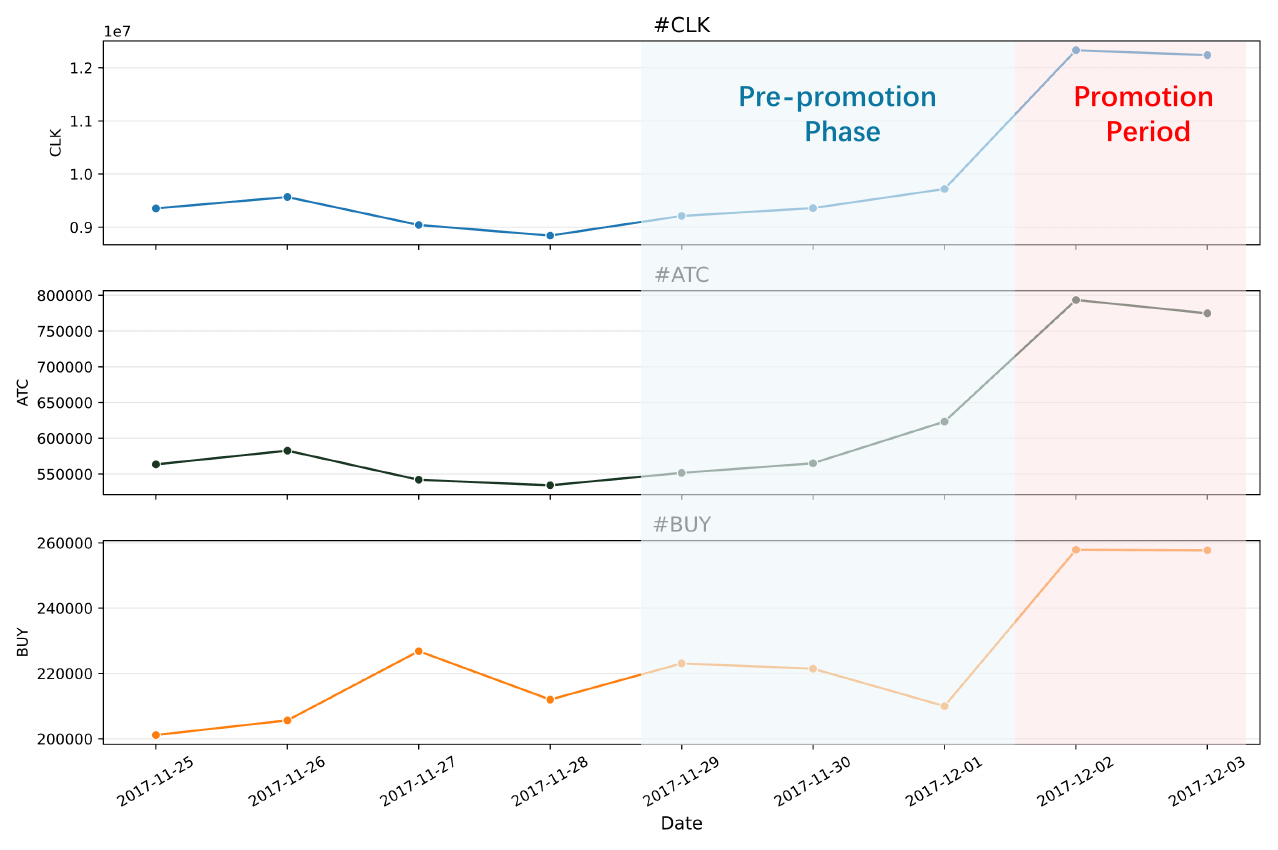}\\
          (a)
      \end{minipage}%
      \hfill
      \begin{minipage}[b]{0.48\textwidth}
          \centering
          \includegraphics[width=\linewidth]{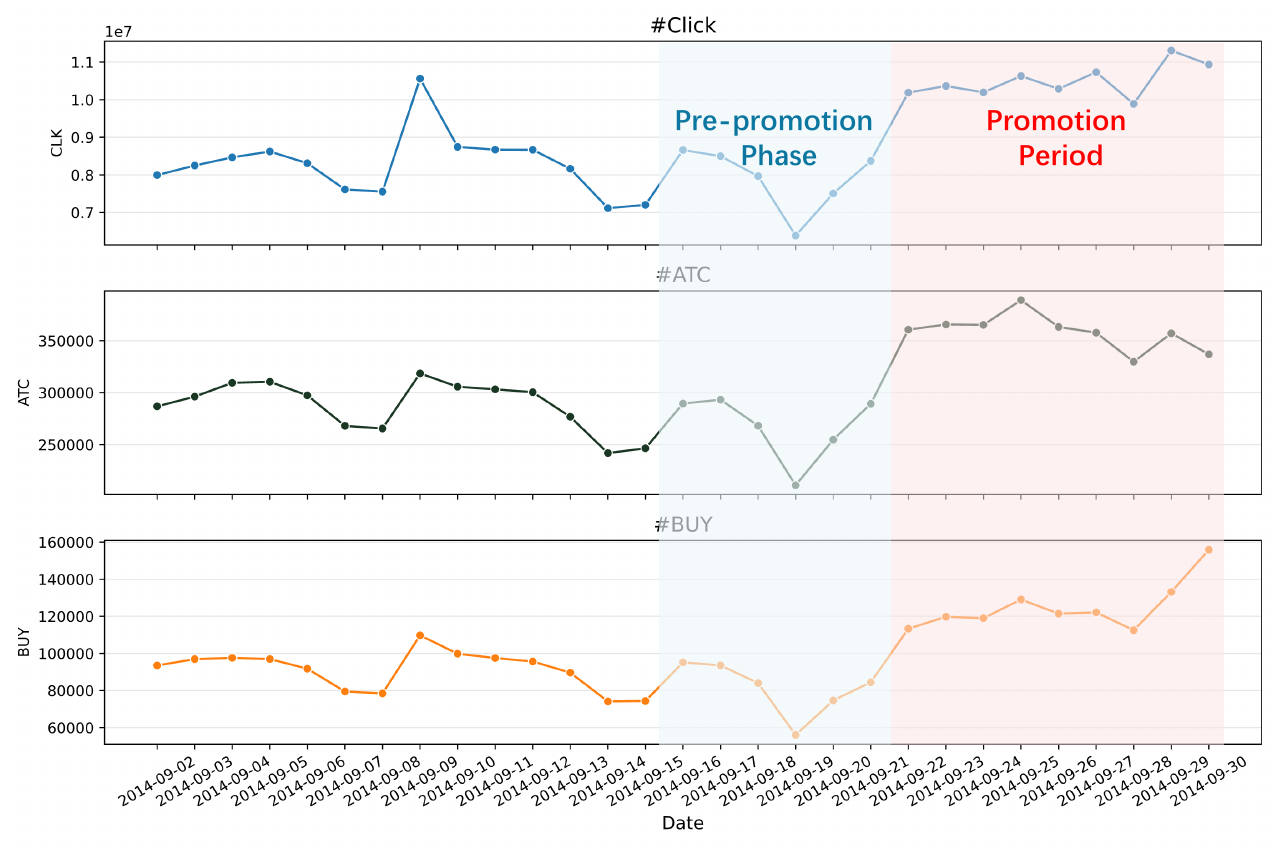}\\
          (b)
      \end{minipage}
      \caption{User action trends for the Taobao (a) and Tmall (b) datasets. "\#CLK" "\#ATC" and "\#BUY" denote the number of clicks, add-to-cart actions, and orders respectively.}
      \label{fig:datasets}
  \end{figure*}

\section{Experiment}
\subsection{Experimental Setting}
\textbf{Dataset.} We use two public datasets, Taobao\footnote{https://tianchi.aliyun.com/dataset/649} and Tmall\footnote{https://tianchi.aliyun.com/dataset/140281}, which record diverse user actions—clicks, cart additions, and purchases along with precise timestamps, enabling accurate calculation of conversion delays in pre-promotion scenarios. 

\textbf{Taobao dataset.} As public datasets lack explicit promotion labels, our exploratory data analysis in Figure \ref{fig:datasets} indicates that the Taobao data spans November 25 to
December 3, 2017. Notably, purchase volume exhibits a declining trend from November 29 to December 1, with concurrent increases in click and
add-to-cart activities, culminating in a sharp spike on December 2. This behavioral pattern is consistent with the promotion-driven user dynamics
described in our Introduction. Therefore we partition the data as follows: November 25–28 serves as regular training data for the pre-trained CVR model,
November 29–December 1 constitutes the pre-promotion window, and December 2–3 represents the core promotion period. Given the absence of prior
promotion cycles in the dataset, we randomly allocate 80\% of these samples as historical promotion data for model training and reserve the
remaining 20\% for evaluation.

\textbf{Tmall dataset.} Following the same rationale, we restrict our analysis of the Tmall dataset to September 2024. Specifically, we designate September 15–20 as the pre-promotion phase and September 21–30 as the promotion period, while utilizing all preceding data to train the regular CVR model. The data
partitioning strategy mirrors that employed for the Taobao dataset, with an identical 80/20 split ratio for historical promotion training and
evaluation.

\textbf{Industrial dataset.} Additionally, we leverage proprietary real-world data from our e-commerce advertising platform, comprising click logs and conversion labels across multiple countries from July 1 to November 11, 2023. Key promotional event days are D7 (July 7), D8 (August 8), D9 (September 9), D10 (October 10), and D11 (November 11). For each event, the three preceding days are defined as the pre-promotion period. The pre-promotion window should be chosen according to the marketing campaign cycle. On our platform, the typical  pre-promotion warm-up period is 3 to 7 days.

All the clicks during pre-promotion phase in all datasets, completed purchase during the promotion day, are labeled as delayed conversions.

\textbf{Metrics} In this paper, We evaluate the ranking performance of our proposed CM-DCM with AUC and Negative log-likelihood (NLL). However, for different ranking tasks, we use different testing sets to calculate the metrics. \textbf{$AUC_{delay}$} assesses the pairwise ranking performance of the classification results between the delayed conversion and non-conversion samples. \textbf{$NLL_{delay}$} is computed with delayed conversion. \textbf{$AUC_{all}$} assesses the pairwise ranking performance of the classification results between the real positive conversion samples and non-conversion negative samples.

\textbf{Compared methods.}\label{sec:baseline} Fisrtly, we trained on 30-day same-day conversion data (no promotion/delayed labels); serving as a base model for fine-tuning. And We selected the following methods as baselines. Delay conversion model FNW \cite{FNW}, ES-DFM \cite{ESC-DFM}, DEFER \cite{DEFER} and DEFUSE \cite{DEFUSE}. For all the methods, there are two strategies to fine-tune them based on \textbf{Pre-trained} CVR model: \textbf{Daily Delayed Strategy}: Train on 20-day continuous data with 1-day delay window, using original duplication/reweighting methods. \textbf{Pre-promotion Reuse Strategy}: Assign delayed conversions in pre-promotion phase to promotion day, applying model-specific duplication and weighting per original designs. And also HDR \cite{HDR}, a conversion modeling method for fluctuating data. 

\textbf{Hyperparameter settings} For offline experiments, we use the Adagrad optimizer (learning rate 0.001) with all hidden layer sizes set to [512, 256, 128], gate MLPs sizes set to [128, 64]. And the batch size is set to 1024. The loss coefficients $\lambda$ and $\lambda_c$ were also determined by a basic parameter search to report the optimal results.

\subsection{Experimental Results}

\begin{table*}[t]
    \centering
    \caption{The overall performance comparison in terms of \(AUC_{all}\), \(AUC_{delay}\) and \(NLL_{delay}\). We run each set of experiments five times and report the average results. $^{\ast}$ indicates p-value < 0.05 in the significance test.}
    \begin{tabular}{l|c|c|c|c|c|c|c|c|c}
         \toprule
         Dataset & \multicolumn{3}{c|}{$Taobao$} & \multicolumn{3}{c|}{$Tmall$} & \multicolumn{3}{c}{$Industrial$}\cr
         \midrule
         Metrics & \(AUC_{all}\) & \(AUC_{delay}\) & \(NLL_{delay}\) & \(AUC_{all}\) & \(AUC_{delay}\) & \(NLL_{delay}\) & \(AUC_{all}\) & \(AUC_{delay}\) & \(NLL_{delay}\) \\
         \midrule
         \midrule
         pre-trained   & 0.7716 & 0.7821 & 0.0357 & 0.6694 & 0.6782 & 0.0371 & 0.7766 & 0.7874  & 0.0328 \\
        FNW  & 0.7737 & 0.7842 & 0.0326 & 0.6730 & 0.6811 & 0.0362 & 0.7822 & 0.7935 & 0.032 \\
        ES-DFM   & 0.7759 & 0.7844 & 0.0322 & 0.6801 & 0.6915 & 0.0370 & 0.7896 & 0.7972 & 0.0314 \\
        DEFER & 0.7807 & 0.7933 & 0.0275 & 0.6993 & 0.7114 & 0.0360 & 0.8101 & 0.8221 & 0.0290 \\
        DEFUSE   & 0.7858 & 0.7905 & 0.0274 & 0.6921 & 0.7034 & 0.0366 & 0.8146 & 0.8244 & 0.0274 \\
        \midrule
        FNW\_{reuse} & 0.8176 & 0.8301 & 0.0268 & 0.7011 & 0.7135 & 0.0353 & 0.8226 & 0.8232 & 0.0263 \\
        ES-DFM\_{reuse} & 0.8205 & 0.8302 & 0.0266 & 0.7099 & 0.7225 & 0.0349 & 0.8233 & 0.8342 & 0.0249\\
        DEFER\_{reuse}  & 0.8258 & 0.8368 & 0.0261 & 0.7235 & 0.7356 & 0.0348 & 0.8252 & 0.8342 & 0.0250\\
        DEFUSE\_{reuse} & 0.8359 & 0.8441 & 0.0260 & 0.7182 & 0.7299 & 0.0338 & 0.8285 & 0.8389 & 0.0261\\
        \midrule
        HDR & 0.8182 & 0.8287 & 0.0270 & 0.7202 & 0.7301  & 0.0350 & 0.8267 & 0.8356 & 0.0262  \\
        \midrule
        \textbf{CM-DCM } & \textbf{0.8588$^{\ast}$} & \textbf{0.8573$^{\ast}$} & \textbf{0.0229$^{\ast}$} & \textbf{0.7401$^{\ast}$} & \textbf{0.7532$^{\ast}$} & \textbf{0.0289$^{\ast}$} & \textbf{0.8570$^{\ast}$} & \textbf{0.8801$^{\ast}$} & \textbf{0.0238$^{\ast}$} \\
         \bottomrule
    \end{tabular}
    \label{tab:perform}
\end{table*}

\begin{table}[t]
    \centering
    \caption{Ablation study on industrial dataset.}
    \begin{tabular}{l|c|c|c}
        \hline
        \hline Method & \(AUC_{all}\) & \(AUC_{delay}\) & \(NLL_{delay}\)  \\
        \hline 
               w/o All-cvr & 0.8411 & 0.8519  & 0.0377 \\
               w/o PG & 0.8510 & 0.8669 & 0.0308  \\
               w/o CM & 0.8512 & 0.8678 & 0.0297 \\
               w/o CCRA & 0.8392 & 08468  & 0.0399 \\
        \hline \textbf{CM-DCM} & \textbf{0.8570} &\textbf{0.8801} &\textbf{0.0238} \\
        \hline
        \hline
    \end{tabular}
    \label{tab:ablation}
    \vspace{-15px}
\end{table}

\textbf{Overall performance.} We run each set of experiments five times and report the average results, along with the statistical significance indicators comparing our method against the best baseline. From Table \ref{tab:perform} we observe that CM-DCM significantly outperforms all baselines in pre-promotion delayed-CVR ranking (AUC/NLL) across countries. First, it surpasses the pre-trained same-day CVR model—highlighting the inadequacy of conventional CVR models for delayed patterns. Second, it beats daily-delay strategies (even with 1-day windows), confirming such methods fail to capture pre-promotion→promotion-day conversion dynamics. Third, it improves upon historical data reuse (HDR) and reuse-enhanced delayed models, owing to: (i) multi-task decoupling of direct/delayed conversion, (ii) personalized gating for adaptive transfer from pre-trained CVR/ATC models, and (iii) explicit counterfactual modeling of ATC→delayed conversion causality—collectively enabling precise distinction between true negatives and promotion-delayed positives.

\textbf{Ablation study.} To verify the effectiveness of such a design, we consider a series of variants of our CM-DCM model for comparison through offline experiments and conduct them on our industrial dateset:
 \textbf{ w/o All-cvr }: This variant removes the transition probability from add-to-cart to delayed cvr module introduced in Sec. \ref{sec:pact}.
 \textbf{ w/o PG}: This variant removes the Personalized Gated module for personalized information transfer from the CVR/ATC model. 
 \textbf{ w/o CM}: This variant removes CM loss
 \textbf{ w/o CCRA}: This variant removes the whole module of counterfactual causal regularization for ATC effect. The result of experiments in Table \ref{tab:ablation} confirmed the effectiveness of each module.

\subsection{Online A/B Experiment}
During the pre-promotion phases of the two largest annual sales promotions, Double 11, and Double 12, We deployed CM-DCM as the experimental group in our industrial e-commerce advertising platform, with the control group using the baseline regular daily CVR model without optimization. Each group covered 20\% of total traffic and the experiment lasted for six days in total. As shown in Table \ref{tab:online_exp}, CM-DCM achieved online revenue increases of +7.87\% and a gain of +4.24\% in delayed GMV, as well as +1.42\% in overall GMV, respectively. This significant improvement demonstrates the effectiveness of our proposed approach. Meanwhile, because the delayed conversion Predictor calculation and the original CVR Predictor run in parallel and the Gating networks are relatively small, the model inference P99 latency increased slightly by 2 ms.

\begin{table}[h]
\centering
\small
\caption{Online Performance of CM-DCM.}
\label{tab:online_exp}
\vspace{-2mm} 
\begin{tabular}{lccccc}
\toprule
Metric & Revenue & Delayed GMV & Overall GMV & P99 Latency\\ 
\midrule
CM-DCM   & +7.87\% & +4.24\% & +1.42\% & +2ms \\
\bottomrule
\end{tabular}
\vspace{-10px}
\end{table}

\section{Conclusion}

In this paper, we propose CM-DCM (Counterfactual Multi-task Delayed Conversion Model) to address the unique delayed conversion challenge in the pre-promotion phase—where conversions are not continuously delayed but sharply concentrated on the promotion day. CM-DCM integrates three key innovations: (i) a multi-task architecture decoupling direct and delayed conversions using historical pre-promotion data; (ii) a counterfactual causal regularizer to model the causal effect of pre-promotion add-to-cart (ATC) on delayed conversion; and (iii) a personalized gating mechanism for adaptive knowledge transfer from pre-trained daily CVR/ATC models. Extensive experiments and online A/B tests confirm that CM-DCM significantly outperform all baseline models.

\bibliographystyle{ACM-Reference-Format}
\balance
\bibliography{ref}

@String{Computing = "Computing" }

@misc{soft_ret,
      title={Soft Retargeting Network for Click Through Rate Prediction}, 
      author={Xiaochen Li and Xin Song and Pengjia Yuan and Xialong Liu and Yu Zhang},
      year={2022},
      eprint={2206.01894},
      archivePrefix={arXiv},
      primaryClass={cs.IR},
      url={https://arxiv.org/abs/2206.01894}, 
}

@inproceedings{LREA,
  title={Lrea: Low-rank efficient attention on modeling long-term user behaviors for ctr prediction},
  author={Song, Xin and Li, Xiaochen and Hu, Jinxin and Wen, Hong and Chen, Zulong and Zhang, Yu and Zeng, Xiaoyi and Zhang, Jing},
  booktitle={Proceedings of the 48th International ACM SIGIR Conference on Research and Development in Information Retrieval},
  location = {Padua, Italy},
  series = {SIGIR '25},
  pages={2843--2847},
  year={2025}
}

@inproceedings{ESMM,
author = {Ma, Xiao and Zhao, Liqin and Huang, Guan and Wang, Zhi and Hu, Zelin and Zhu, Xiaoqiang and Gai, Kun},
title = {Entire Space Multi-Task Model: An Effective Approach for Estimating Post-Click Conversion Rate},
year = {2018},
isbn = {9781450356572},
publisher = {Association for Computing Machinery},
address = {New York, NY, USA},
url = {https://doi.org/10.1145/3209978.3210104},
doi = {10.1145/3209978.3210104},
booktitle = {The 41st International ACM SIGIR Conference on Research \& Development in Information Retrieval},
pages = {1137–1140},
numpages = {4},
location = {Ann Arbor, MI, USA},
series = {SIGIR '18}
}

@inproceedings{ESCM2,
author = {Wang, Hao and Chang, Tai-Wei and Liu, Tianqiao and Huang, Jianmin and Chen, Zhichao and Yu, Chao and Li, Ruopeng and Chu, Wei},
title = {ESCM2: Entire Space Counterfactual Multi-Task Model for Post-Click Conversion Rate Estimation},
year = {2022},
isbn = {9781450387323},
publisher = {Association for Computing Machinery},
address = {New York, NY, USA},
url = {https://doi.org/10.1145/3477495.3531972},
doi = {10.1145/3477495.3531972},
booktitle = {Proceedings of the 45th International ACM SIGIR Conference on Research and Development in Information Retrieval},
pages = {363–372},
numpages = {10},
location = {Madrid, Spain},
series = {SIGIR '22}
}

@inproceedings{ESMMBD,
author = {Wen, Hong and Zhang, Jing and Wang, Yuan and Lv, Fuyu and Bao, Wentian and Lin, Quan and Yang, Keping},
title = {Entire Space Multi-Task Modeling via Post-Click Behavior Decomposition for Conversion Rate Prediction},
year = {2020},
isbn = {9781450380164},
publisher = {Association for Computing Machinery},
address = {New York, NY, USA},
url = {https://doi.org/10.1145/3397271.3401443},
doi = {10.1145/3397271.3401443},
booktitle = {Proceedings of the 43rd International ACM SIGIR Conference on Research and Development in Information Retrieval},
pages = {2377–2386},
numpages = {10},
location = {Virtual Event, China},
series = {SIGIR '20}
}

@article{ModelingTR,
  title={Modeling Task Relationships in Multi-task Learning with Multi-gate Mixture-of-Experts},
  author={Jiaqi Ma and Zhe Zhao and Xinyang Yi and Jilin Chen and Lichan Hong and Ed H. Chi},
  journal={Proceedings of the 24th ACM SIGKDD International Conference on Knowledge Discovery \& Data Mining},
  year={2018},
  url={https://api.semanticscholar.org/CorpusID:50770252}
}

@article{ESDFM, 
title={Capturing Delayed Feedback in Conversion Rate Prediction via Elapsed-Time Sampling}, 
volume={35}, 
url={https://ojs.aaai.org/index.php/AAAI/article/view/16587}, 
DOI={10.1609/aaai.v35i5.16587} , 
number={5}, 
journal={Proceedings of the AAAI Conference on Artificial Intelligence}, 
author={Yang, Jia-Qi and Li, Xiang and Han, Shuguang and Zhuang, Tao and Zhan, De-Chuan and Zeng, Xiaoyi and Tong, Bin}, 
year={2021}, month={May}, pages={4582-4589} }

@inproceedings{DEFER,
author = {Gu, Siyu and Sheng, Xiang-Rong and Fan, Ying and Zhou, Guorui and Zhu, Xiaoqiang},
title = {Real Negatives Matter: Continuous Training with Real Negatives for Delayed Feedback Modeling},
year = {2021},
isbn = {9781450383325},
publisher = {Association for Computing Machinery},
address = {New York, NY, USA},
url = {https://doi.org/10.1145/3447548.3467086},
doi = {10.1145/3447548.3467086},
booktitle = {Proceedings of the 27th ACM SIGKDD Conference on Knowledge Discovery \& Data Mining},
pages = {2890–2898},
numpages = {9},
location = {Virtual Event, Singapore},
series = {KDD '21}
}

@inproceedings{FNW,
author = {Ktena, Sofia Ira and Tejani, Alykhan and Theis, Lucas and Myana, Pranay Kumar and Dilipkumar, Deepak and Husz\'{a}r, Ferenc and Yoo, Steven and Shi, Wenzhe},
title = {Addressing Delayed Feedback for Continuous Training ith Neural Networks in CTR Prediction},
year = {2019},
isbn = {9781450362436},
publisher = {Association for Computing Machinery},
address = {New York, NY, USA},
url = {https://doi.org/10.1145/3298689.3347002},
doi = {10.1145/3298689.3347002},
booktitle = {Proceedings of the 13th ACM Conference on Recommender Systems},
pages = {187–195},
numpages = {9},
location = {Copenhagen, Denmark},
series = {RecSys '19}
}

@inproceedings{DFM,
author = {Chapelle, Olivier},
title = {Modeling delayed feedback in display advertising},
year = {2014},
isbn = {9781450329569},
publisher = {Association for Computing Machinery},
address = {New York, NY, USA},
url = {https://doi.org/10.1145/2623330.2623634},
doi = {10.1145/2623330.2623634},
booktitle = {Proceedings of the 20th ACM SIGKDD International Conference on Knowledge Discovery and Data Mining},
pages = {1097–1105},
numpages = {9},
location = {New York, New York, USA},
series = {KDD '14}
}

@inproceedings{ESC-DFM,
author = {Zhao, Yunfeng and Yan, Xu and Gui, Xiaoqiang and Han, Shuguang and Sheng, Xiang-Rong and Yu, Guoxian and Chen, Jufeng and Xu, Zhao and Zheng, Bo},
title = {Entire Space Cascade Delayed Feedback Modeling for Effective Conversion Rate Prediction},
year = {2023},
isbn = {9798400701245},
publisher = {Association for Computing Machinery},
address = {New York, NY, USA},
url = {https://doi.org/10.1145/3583780.3615475},
doi = {10.1145/3583780.3615475},
booktitle = {Proceedings of the 32nd ACM International Conference on Information and Knowledge Management},
pages = {4981–4987},
numpages = {7},
location = {Birmingham, United Kingdom},
series = {CIKM '23}
}

@inproceedings{chang2023pepnet,
author = {Chang, Jianxin and Zhang, Chenbin and Hui, Yiqun and Leng, Dewei and Niu, Yanan and Song, Yang and Gai, Kun},
title = {PEPNet: Parameter and Embedding Personalized Network for Infusing with Personalized Prior Information},
year = {2023},
isbn = {9798400701030},
publisher = {Association for Computing Machinery},
address = {New York, NY, USA},
url = {https://doi.org/10.1145/3580305.3599884},
doi = {10.1145/3580305.3599884},
booktitle = {Proceedings of the 29th ACM SIGKDD Conference on Knowledge Discovery and Data Mining},
pages = {3795–3804},
numpages = {10},
location = {Long Beach, CA, USA},
series = {KDD '23}
}

@inproceedings{DEFUSE,
author = {Chen, Yu and Jin, Jiaqi and Zhao, Hui and Wang, Pengjie and Liu, Guojun and Xu, Jian and Zheng, Bo},
title = {Asymptotically Unbiased Estimation for Delayed Feedback Modeling via Label Correction},
year = {2022},
isbn = {9781450390965},
publisher = {Association for Computing Machinery},
address = {New York, NY, USA},
url = {https://doi.org/10.1145/3485447.3511965},
doi = {10.1145/3485447.3511965},
booktitle = {Proceedings of the ACM Web Conference 2022},
pages = {369–379},
numpages = {11},
location = {Virtual Event, Lyon, France},
series = {WWW '22}
}

@inproceedings{HDR,
author = {Chan, Zhangming and Zhang, Yu and Han, Shuguang and Bai, Yong and Sheng, Xiang-Rong and Lou, Siyuan and Hu, Jiacen and Liu, Baolin and Jiang, Yuning and Xu, Jian and Zheng, Bo},
title = {Capturing Conversion Rate Fluctuation during Sales Promotions: A Novel Historical Data Reuse Approach},
year = {2023},
isbn = {9798400701030},
publisher = {Association for Computing Machinery},
address = {New York, NY, USA},
url = {https://doi.org/10.1145/3580305.3599788},
doi = {10.1145/3580305.3599788},
booktitle = {Proceedings of the 29th ACM SIGKDD Conference on Knowledge Discovery and Data Mining},
pages = {3774–3784},
numpages = {11},
location = {Long Beach, CA, USA},
series = {KDD '23}
}

@article{effects_promption,
  title={The effects of sales promotion strategy, product appeal and consumer traits on reminder impulse buying behaviour},
  author={Liao, Shu-Ling and Shen, Yung-Cheng and Chu, Chia-Hsien},
  journal={International Journal of Consumer Studies},
  volume={33},
  number={3},
  pages={274--284},
  year={2009},
  publisher={Wiley Online Library}
}

@article{inf_prootion,
  title={Analyzing the influence of sales promotion on customer purchasing behavior},
  author={Familmaleki, Mahsa and Aghighi, Alireza and Hamidi, Kambiz},
  journal={International Journal of Economics \& management sciences},
  volume={4},
  number={4},
  pages={1--6},
  year={2015}
}

@article{post-promotion,
title = {The effect of sales promotion on post-promotion brand preference: A meta-analysis},
journal = {Journal of Retailing},
volume = {82},
number = {3},
pages = {203-213},
year = {2006},
issn = {0022-4359},
doi = {https://doi.org/10.1016/j.jretai.2005.10.001},
url = {https://www.sciencedirect.com/science/article/pii/S0022435906000388},
author = {Devon DelVecchio and David H. Henard and Traci H. Freling}
}

@inproceedings{PEPNet,
author = {Chang, Jianxin and Zhang, Chenbin and Hui, Yiqun and Leng, Dewei and Niu, Yanan and Song, Yang and Gai, Kun},
title = {PEPNet: Parameter and Embedding Personalized Network for Infusing with Personalized Prior Information},
year = {2023},
isbn = {9798400701030},
publisher = {Association for Computing Machinery},
address = {New York, NY, USA},
url = {https://doi.org/10.1145/3580305.3599884},
doi = {10.1145/3580305.3599884},
booktitle = {Proceedings of the 29th ACM SIGKDD Conference on Knowledge Discovery and Data Mining},
pages = {3795–3804},
numpages = {10},
location = {Long Beach, CA, USA},
series = {KDD '23}
}
\end{document}